\documentclass[letterpaper,12pt]{article}   

\usepackage{arxiv} 

\begin{document}
\title{Twisted split-ring-resonator photonic metamaterial with huge optical activity}

\author{M. Decker$^{1}$, R. Zhao$^{2,3}$, C.\,M. Soukoulis$^{3,4}$, S. Linden$^{1}$, and M. Wegener$^{1}$ }

\affiliation{$^1$Institut f\"ur Angewandte Physik and DFG-Center for Functional Nanostructures (CFN), Karlsruhe Institute of Technology (KIT), D-76128 Karlsruhe, Germany, and \\Institut f\"ur Nanotechnologie, Karlsruhe Institute of Technology (KIT), D-76021 Karlsruhe, Germany}
\affiliation{$^2$Applied Optics Beijing Area Major Laboratory, Department of Physics, Beijing Normal University, Beijing 100875, China}
\affiliation{$^3$Ames Laboratory and Department of Physics and Astronomy, Iowa State University, Ames, Iowa 50011, U.S.A.}
\affiliation{$^4$IESL-FORTH and Dept. of Materials Science  and Technology, University of Crete, 71110, Heraklion, Greece}

\begin{abstract}
\noindent 
Coupled split-ring-resonator metamaterials have previously been shown to exhibit large coupling effects, which are a prerequisite for obtaining large effective optical activity. By a suitable lateral arrangement of these building blocks, we completely eliminate linear birefringence and obtain pure optical activity and connected circular optical dichroism. Experiments at around 100-THz frequency and corresponding modeling are in good agreement. Rotation angles of about 30 degrees for 205\,nm sample thickness are derived.\\
\end{abstract}

\maketitle

\newpage

Optical activity in effective media refers to a difference $\Delta n=n_{\rm RCP}-n_{\rm LCP}$ in the real parts of the refractive indices for left- and right handed circularly polarized incident light. The Kramers-Kronig relations connect these differences to the imaginary parts of the refractive indices, i.e., to circular dichroism. Optical activity requires a magnetic-dipole response mediated by the electric field of the electromagnetic light wave \cite{Plum2009,Zhang2009,Wegener2009}. In natural substances like solutions of chiral molecules, these effects are quite small, i.e., $|\Delta n| \ll 1$. Strong effective magnetic dipoles can arise from the coupling of Mie-like electric-dipole resonances. For example, coupled gold crosses have recently been discussed in this context \cite{Decker2009,Zhou2009}. Yet stronger coupling effects have been reported for twisted split-ring resonators (SRR) \cite{Liu2009}. However, the latter structures do not only exhibit strong optical activity but also strong linear birefringence. A clean separation is desirable for both understanding the physics and for applications such as ``poor-man's'' optical isolators \cite{Thiel2007}.

In this Letter, we discuss a novel lateral arrangement of such twisted SRR that, by symmetry, exhibits vanishing linear optical birefringence.
This structure has been mentioned theoretically previously \cite{Liu2008}. The circular dichroism measured by us translates to values of $|\Delta n|\approx 2$, which outperform our previous best results on twisted crosses \cite{Decker2009} by about a factor of six.

\begin{figure}[Hb]
\centerline{\includegraphics[width=8.255cm,keepaspectratio]{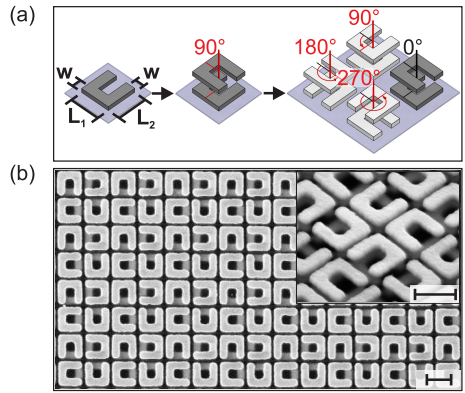}}
\caption{(a) Illustration of our metamaterial's chiral unit cell composed of gold SRR. The lateral dimensions of the SRR are indicated on the left-hand side. (b) Electron micrographs of a typical fabricated structure. The normal-incidence image illustrates the high alignment accuracy of the two stacked layers. The inset shows an oblique view onto the sample. The scalebars are $400\rm nm$.}
\end{figure}

The twisted SRR pair in Ref.\,\cite{Liu2009} exhibits an obvious asymmetry between the $x$ and the $y$-direction. To avoid the resulting linear birefringence, we design a square unit cell which is formed by four of these pairs, whereby the pairs are rotated by 0, 90, 180, and 270 degrees with respect to the stacking axis (see Fig.\,1(a)). The resulting overall crystal structure has four-fold rotational symmetry, no center of inversion, and no mirror planes. Hence, it is truly chiral. We emphasize that the detailed arrangement of the SRR does matter. For example, the structure recently published in Ref.\,\cite{Xiong2009} exhibits linear birefringence in contrast to ours. The different symmetry of the structure in Ref.\,\cite{Xiong2009} can be compared to the symmetry of the structure recently published by us \cite{Decker2009b,Decker2009c} exhibiting linear eigenpolarizations of +45 degrees and -45 degrees. Indeed, this also holds true for the arrangement of SRR in Ref.\,\cite{Xiong2009}. Our theoretical calculations for their structure (not shown) also reveal only extremely small optical activity and circular dichroism. This finding highlights the importance of the relative arrangement of the SRR pairs for the optical response.

Fabrication of the two-layer chiral medium shown in Fig.\,1 requires advanced nanofabrication, i.e., two successive electron-beam-lithography steps and an intermediate planarization process via a spin-on dielectric \cite{Subramania2004}. Starting with the first functional layer written by electron-beam lithography, we planarize the sample via a 500-nm thick spacer layer of commercially available spin-on dielectric (IC1-200, {\it Futurrex Inc.}) and a subsequent thinning via reactive-ion etching ($\rm SF_6$, {\it Plasmalab80Plus, Oxford Instruments Plasma Technology}). Next, we process the second functional layer of split-ring resonators via another electron-beam-lithography step carefully aligned relative to the first layer via alignment markers. As a result, we achieve an alignment mismatch of the first and the second layer of below $10 \rm nm$ over the entire sample footprint of $100 {\rm  \mu m} \times 100 {\rm  \mu m}$. Electron micrographs of a fabricated structure are shown in Fig.\,1(b). The two functional layers are separated by a 85-nm thick spacer layer. All samples are fabricated on a glass substrate covered with a 5-nm thin film of indium-tin-oxide (ITO). Obviously, the sample quality is very high. In particular, no misalignment between the two split-ring-resonators in each pair is detectable. The in-plane lattice constant of the set of four SRR pairs of $a_{xy}=a=885\,\rm nm$ is significantly smaller than the resonance wavelengths of about 3\,$\rm \mu m$. 

\begin{figure}[Hb]
\centerline{\includegraphics[width=8.255cm,keepaspectratio]{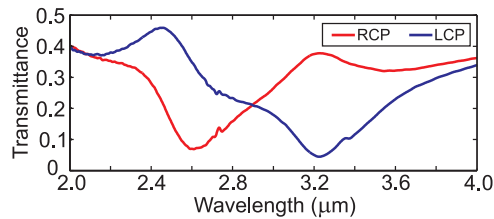}}
\caption{Measured normal-incidence intensity transmittance spectra for left-handed circularly polarized (LCP) and right-handed circularly polarized (RCP) incident light onto the sample shown in Fig.\,1(b).}
\end{figure}

For optical characterization, we use a commercial Fourier-transform microscope spectrometer ({\it Bruker Tensor 27} with {\it Bruker Hyperion 1000}) combined with a linear $\rm CaF_2$ ``High Extinction Ratio'' holographic polarizer and a super-achromatic $\rm MgF_2$ based quarter-wave plate ({\it Bernhard Halle Nachfl.}, 2.5 to 7.0 $\rm \mu m$) that can be rotated from the outside of the microscope \cite{Gansel2009}. Furthermore, we have modified the reflective $\times 36$ Cassegrain lens (NA = 0.5) by introducing a small diaphragm such that the full opening angle of the light incident onto the sample is reduced to about five degrees. The sample is tilted such that we achieve actual normal incidence of light onto the sample. Normalization of the transmittance spectra is with respect to the transmittance of the glass substrate, the ITO- and the spacer layer. 

The measured transmittance spectra (see Fig.\,2) for left-handed circularly polarized (LCP) and right-handed circularly polarized (RCP) light are obviously quite different. Indeed, the observed effects are much stronger compared to our ``planar chiral'' metamaterial structure \cite{Decker2007} as well as compared to our twisted-cross metamaterial structure\cite{Decker2009}. Precisely, the circular dichroism, i.e., the difference between RCP and LCP transmittance, reaches values as large as 33\,\% for the present two-layer twisted-SRR metamaterial. 

\begin{figure}[Hb]
\centerline{\includegraphics[width=8.255cm,keepaspectratio]{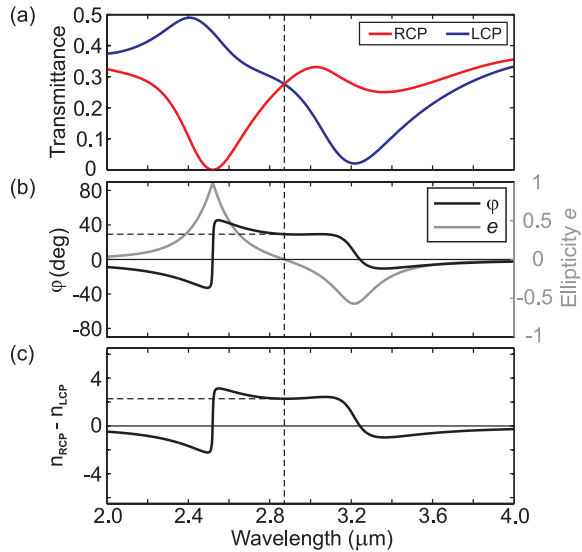}}
\caption{Calculated optical properties. (a) Normal-incidence transmittance spectra that can be directly compared to the experiment shown in Fig.\,2. The intensity conversion (not shown) is below $10^{-5}$ for the entire spectral range. (b) Calculated rotation angle $\varphi$ and tangent $e$ of the ellipticity angle of the transmitted light for linearly polarized incident light. (c) Difference of refractive indices $\Delta n=n_{\rm RCP}-n_{\rm LCP}$ retrieved from the complex transmittance and reflectance spectra.}
\end{figure}

Unfortunately, our experimental setup does not allow for analyzing the emerging polarization of light. To investigate this aspect and to understand the nature of the observed resonances, we perform additional numerical modeling. We use a finite-element frequency-domain approach provided by Comsol MultiPhysics supported by finite-integral time-domain calculations by CST MicroWave Studio. The lateral geometrical parameters of the SRR are $L_1=380\,\rm nm$, $L_2=350\,\rm nm$, and $w=115\,\rm nm$ (see Fig.\,1(a)). The gold thickness in each layer is $60\,\rm nm$, that of the spacer layer $85\,\rm nm$. The unit cell is arranged in a square lattice with an in-plane lattice constant of $a=885\,\rm nm$. The gold optical properties are modeled by a free-electron Drude model with plasma frequency $\omega_{\rm pl}=2\pi \times 2133\,\rm THz$ and collision frequency $\omega_{\rm coll}=2\pi\times 33\,\rm THz$. The refractive indices of the glass substrate and the spin-on dielectric are 1.45 and 1.41, respectively. The thin ITO film is neglected. 

The calculated results in Fig.\,3 (a) nicely qualitatively agree with our experimental findings in Fig.\,2. Furthermore, we find very little intensity conversion (below $10^{-5}$) of circular polarization throughout the entire spectral range, i.e., LCP (RCP) incident light emerges as LCP (RCP) transmitted light. This means that LCP and RCP are very nearly the eigen-polarizations of the Jones Matrix of our chiral metamaterial structure. Fig.\,3(b) shows the corresponding calculated rotation angle of linearly polarized incident light, that is the effects of optical activity. This part of the figure also shows the calculated tangent, $e=tan(\eta)$, of the ellipticity angle $\eta$, i.e., the ratio between the semiminor and the semimajor axis of the polarization ellipse. $e=0$ corresponds to linear polarization, $e=\pm 1$ to circular polarization. Obviously, the rotation angle $\varphi$ in Fig.\,3(b) exhibits a resonance behavior. In resonance, the ellipticity approaches $e=1$. For pure optical activity we need $e=0$. At this zero crossing in Fig.\,3 (dashed line), we find a rotation angle of about 30 degrees for a metamaterial thickness of just $205 {\rm nm}$. Employing the usual parameter retrieval \cite{Kwon2008} accounting for the glass substrate leads to the difference between RCP and LCP refractive indices $\Delta n=n_{\rm RCP}-n_{\rm LCP}$ shown in Fig.\,3(c). Values of $|\Delta n|\approx 2$ are found. As expected, the spectral shape of the retrieved index difference $\Delta n$ closely resembles the rotation angle (Fig.\,3(b)) that is directly obtained from the calculated transmission phases.

\begin{figure}[Hb]
\centerline{\includegraphics[width=8.255cm,keepaspectratio]{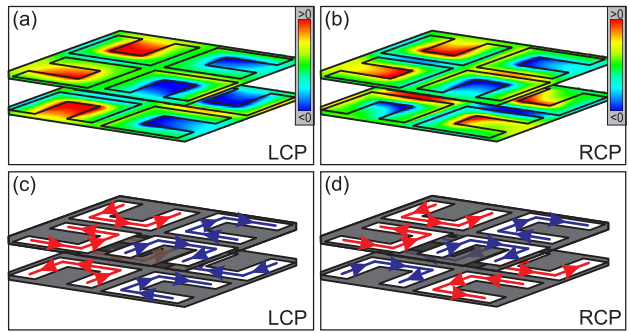}}
\caption{False-color plots of the axial component of the local magnetic field in two planes cutting through the SRR layers for (a) LCP incidence at $3.2\rm \mu m$ wavelength and (b) for RCP incidence at $2.5\rm \mu m$. Schemes of the corresponding underlying electric currents within the SRR are shown in (c) and (d), respectively.}
\end{figure}
 
Finally, we study the nature of the resonances seen in Figs.\,2 and 3. The calculated axial component of the local magnetic field is illustrated in Fig.\,4. For the low-frequency LCP resonance, the magnetic moments within each SRR pair are obviously parallel. In contrast, for the high-frequency RCP resonance, they are antiparallel. Clearly, the coupling between the two SRR in each pair is crucial for the observed optical activity. Without SRR coupling, the frequency splitting (or hybridization or avoided crossing) between the two modes of the coupled system would become zero and optical activity would disappear.

In conclusion, we have designed, fabricated, and characterized a novel layer-by-layer chiral metamaterial structure operating at optical frequencies that largely outperforms previous designs regarding the achieved level of optical activity. At around 3-$\mu m$ wavelength, we derive rotation angles as large as 30 degrees, corresponding to refractive-index differences of about $|\Delta n|\approx 2$. While completing our manuscript, we have become aware of an independent work on an archive server \cite{archive} presenting a closely related structure, albeit operating at a much larger wavelength of $40-100\,\rm \mu m$ in their experiments. 

We acknowledge support by the European Commission via the project PHOME and by the Bundesministerium f\"ur Bildung und Forschung via the project METAMAT. The research of S.L. is supported through a Helmholtz-Hochschul-Nachwuchsgruppe (VH-NG-232). The PhD education of M.D. is embedded in the Karlsruhe School of Optics \& Photonics (KSOP). Work at Ames Lab was supported by Dept. of Energy (Basic Energy Sciences), contract No. DE-AC02-07CH11358.


\end{document}